\documentclass[lettersize,journal]{IEEEtran}
\usepackage{amsmath,amsfonts}
\usepackage{algorithm}
\usepackage{algpseudocode} 
\usepackage{array} 
\usepackage{color}
\usepackage{textcomp}
\usepackage{stfloats}
\usepackage{url}
\usepackage{verbatim}
\usepackage{graphics} 
\usepackage{epsfig}   
\usepackage{subfigure}
\usepackage{booktabs}   
\usepackage{multirow}
\usepackage{cite}
\newtheorem{theorem}{\textbf{Theorem}}

\def\mathbi#1{\textbf{\em #1}}
   
\begin{document}

\title{Robust Predictive Routing for Internet of Vehicles Leveraging Both V2I and V2V Links}

\author{Yawen Chang,~\IEEEmembership{Student Member,~IEEE} and Xudong Wang,~\IEEEmembership{Fellow,~IEEE}
\thanks{Y. Chang is with the UM-SJTU Joint Institute, Shanghai Jiao Tong University. X. Wang is with the Hong Kong University of Science and Technology (Guangzhou). Corresponding author: Xudong Wang; E-mail: wxudong@ieee.org.}}



\maketitle

\begin{abstract}
With the developments of the Internet of Vehicles (IoV) from 4G to 5G, vehicle-to-infrastructure (V2I) communications are becoming attractive for vehicle users (VUEs) to obtain diverse cloud service through base stations (BSs). To tackle V2I link deterioration caused by blockage and out-of-coverage cases, multi-hop V2X routing with both vehicle-to-vehicle (V2V) and V2I links needs to be investigated. 
However, traditional routing reacts to statistical or real-time information, which may suffer link degradation during path switchover in fast-changing vehicular networks. Predictive routing protocols take timely actions by forecasting link connectivity, but they fail to satisfy specific QoS requirements. Low robustness to link failures is also incurred without considering imperfect prediction.
To build continual paths between VUEs and BSs for QoS provision of cloud service, a robust predictive routing framework (ROPE) is proposed with three major components: 1) an early warning scheme detects V2I link deterioration in advance via predicting vehicle mobility and link signal strength to facilitate seamless path switchover; 2) a virtual routing mechanism finds top3 paths that have the highest path strength and satisfy the connectivity and hop count constraints based on the prediction results to fulfill QoS requirements of cloud service; 3) a path verification protocol checks availability and quality of the top3 paths shortly before switchover and activates one qualified path for switchover to ensure routing robustness. We implement ROPE in a simulation framework incorporating real-world urban maps, microscopic traffic generation, geometry-based channel modeling, and offline data analysis as well as online inference. Extensive simulations demonstrate the superiority of ROPE over direct V2I communications and a connectivity-based predictive routing protocol under various scenarios.
\end{abstract}


\section{Introduction}
\IEEEPARstart{T}{he} Internet of Vehicles (IoV), which enables vehicle-to-infrastructure~(V2I), vehicle-to-vehicle~(V2V), and generally vehicle-to-everything~(V2X) communications, has emerged as a key technology for intelligent transportation system (ITS) \cite{lyu2018intelligent}. To fulfill the vision of IoV, 3GPP proposes the first cellular V2X~(C-V2X) standard in 4G and evolves it in 5G, where V2I link is supported by uplink with Uu interface and V2V link is supported by sidelink with PC5 interface \cite{37.985}.

\begin{figure}[t]
	\centering
	\includegraphics[width=\columnwidth]{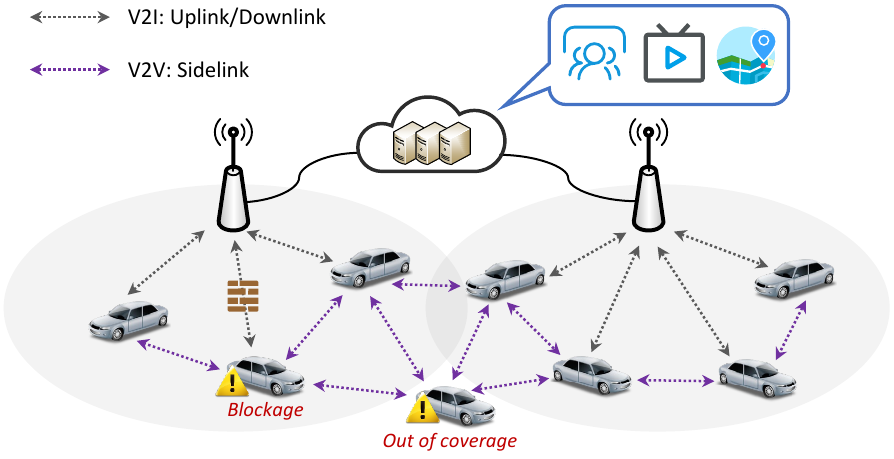}
	\caption{Scenarios of V2I link deterioration.}
	\label{fig-scene}
\end{figure}

With the advancement of autonomous driving, people’s hands will be liberated from the steering wheel and more infotainment demand will come to the fore.
V2I communications enable vehicle users (VUE) to communicate with the cloud server via infrastructure such as base stations (BSs). Therefore, diverse cloud service (e.g., cloud meetings, live streaming, and high-definition maps) can be provided to drivers and passengers for infotainment. 
However, V2I link may suffer deterioration in the following scenarios, which is shown in Fig.~\ref{fig-scene}.
First, blockage due to buildings and foliage is common in urban scenes, which may cause V2I link to encounter intermittence and fail to satisfy QoS requirements of cloud service. 
Second, when vehicles move out of coverage of BSs, the connection between vehicles and BSs is broken, so the cloud service cannot be delivered through V2I links.
To tackle these issues, a multi-hop V2X paradigm leveraging both V2V and V2I links to form a continual path from VUE to BS is necessary and thereby investigated. 
The multi-hop path should have high received signal strength (RSS), high connection duration, and small hop count for satisfying QoS requirements.
Also, VUE should timely switch to the indirect path so that high-quality cloud service can be constantly delivered.

Towards achieving the above goals, there exist three main challenges. 
First is \textit{seamless path switchover}. Rapid speed of vehicles leads to fast link variation and sudden link deterioration. If indirect paths are found and activated after V2I link deterioration is detected, the delivery quality of cloud service will decline during the routing and switching process. Therefore, V2I link deterioration needs to be detected in advance rather than in time. However, V2I link strength is impacted by complex factors such as spatial location and traffic density~\cite{gozalvez2012ieee,granda2020spatial}, which is hard to estimate for deterioration detection.
Second is \textit{QoS-driven path selection}. 
On the one hand, a multi-hop path contains both V2V and V2I links, which have heterogeneous patterns and need to be unified. 
On the other hand, link and path metrics in terms of QoS need to be proactively evaluated for selecting paths that satisfy QoS requirements.
Third is \textit{robustness to link failures}. As the routing decision is made based on estimation or prediction of link status, link failures are likely to happen due to imperfect prediction or abrupt events. To ensure routing robustness, the selected paths need to be verified shortly before switchover and remedy solutions should be quickly launched in case link failures occur.

Yet, none of the existing studies solve these challenging issues.
Reactive methods make online routing decisions based on statistical or real-time information~\cite{38.836,antcolony,onlineseq,peakaverage}, which cannot facilitate seamless path switchover in fast-changing vehicular networks. End-to-end paths may suffer deterioration during switchover, so QoS requirements cannot be satisfied constantly.
Predictive routing has potential to realize seamless path switchover and constant QoS provision for its forecasting ability, but existing work merely focuses on optimizing path connectivity or connection duration \cite{eiza2013evolving,khan2019unsupervised,kumbhar2021novel,linsalata2022map}.
In VoEG \cite{eiza2013evolving} and CVoEG \cite{khan2019unsupervised}, evolving graph theories are leveraged to predict link connection probability as reliability. The Dijkstra algorithm is modified to find the most reliable path among all paths.
In CAR~\cite{kumbhar2021novel}, link duration time is predicted via an analytical model or several machine learning models.
All possible paths between a source and a destination are evaluated using a maximum–minimum approach, which expects that the communication pair stays connected for a longer period. 
In~\cite{linsalata2022map}, cooperation between vehicles and environment is utilized to predict a dynamic line-of-sight (LOS) map. Based on the LOS map, a single-hop relay finding scheme is proposed to maximize network connectivity, so that the impairment of blockage on mmWave vehicular networks in urban scenes can be mitigated.
However, all these predictive methods fail to satisfy specific QoS requirements since they do not make full use of prediction capabilities. Furthermore, remedy solutions to imperfect prediction are not considered, so they bear the limitation of low robustness to link failures.

Though faced with difficulties of fast topology change, vehicular networks are also given opportunities.  
First, different from traditional users, vehicles travel along roads and follow traffic regulations, whose mobility is more predictable and thus can be foreseen based on their past trajectories and actions \cite{moda}. 
Second, with plentiful BSs deployed in urban and storage capacities enhanced in cloud computing, big data of V2X communications can be easily collected through BSs and processed by the cloud server \cite{cloudcomp}.
Given these chances, a robust predictive routing framework (ROPE) is proposed to build a continual path between VUEs and BSs for QoS provision of the cloud service, which addresses all the challenges mentioned earlier. 

The ROPE framework consists of three main components: an early warning scheme, a virtual routing mechanism, and a path verification protocol. 
In the early warning scheme, a context-aware probability neural network (CAPNet) is developed to infer V2I link strength based on predicted mobility of VUEs and traffic density around VUEs. Once the inferred strength is below a configured threshold or the predicted location is outside the coverage range of BSs, early warning is triggered to facilitate routing in advance and seamless path switchover.
In the virtual routing mechanism, V2X virtual topology is formed based on inferred strength of V2I and V2V links as well as predicted vehicle mobility. Link and path metrics in terms of strength, connectivity, and hop count are designed to unify heterogeneous V2X links and make full use of prediction capability on the cloud side.
Given these unified metrics on the virtual topology, a top3 routing algorithm~(TORA) is devised to find three paths between a warned VUE and BS that have the highest path strength while satisfying the connectivity and hop count constraints, so QoS-driven path selection is enabled. The optimization problems are proven to be NP-complete and TORA is analyzed to obtain polynomial complexity. 
For enhancing routing robustness to link failures, a path verification protocol is executed to check actual quality and availability of the top3 paths shortly before switchover. When none of these paths are deemed qualified, a path mending procedure is carried out to generate a qualified path out of two unqualified paths. This qualified path is activated between VUE and BS after switchover so that cloud service with QoS satisfaction can be constantly delivered.

To summarize, the contributions of this paper are as follows:
\begin{itemize}
	\item A robust predictive routing method (ROPE) is proposed to find a continual path between VUE and BS for QoS provision of cloud service.
	\item An early warning scheme is developed to proactively detect V2I link deterioration with the aid of a context-aware probability neural network (CAPNet), which considers both spatial location and traffic density for inferring received signal strength.
	\item A virtual routing mechanism is presented based on unified metrics of strength, connectivity, and hop count, where a top3 routing algorithm solves NP-complete optimization problems under acceptable complexity and generates three best feasible paths.
	\item A path verification protocol is devised to check availability and quality of the three paths shortly before switchover for enhancing robustness, where a path mending procedure is launched to handle link failures.
\end{itemize}

The rest of this paper is organized as follows.
The overall framework of ROPE is presented in Section~\ref{sec-framework}. The early warning scheme is explained in Section~\ref{sec-warning}. The virtual routing mechanism is elaborated in Section~\ref{sec-routing}. The path verification protocol is described in Section~\ref{sec-verification}. Performance evaluation is conducted in Section~\ref{sec-evaluation}. Finally, conclusion and future work are provided in Section~\ref{sec-conclusion}.

\begin{figure}[t]
	\centering
	\includegraphics[width=\columnwidth]{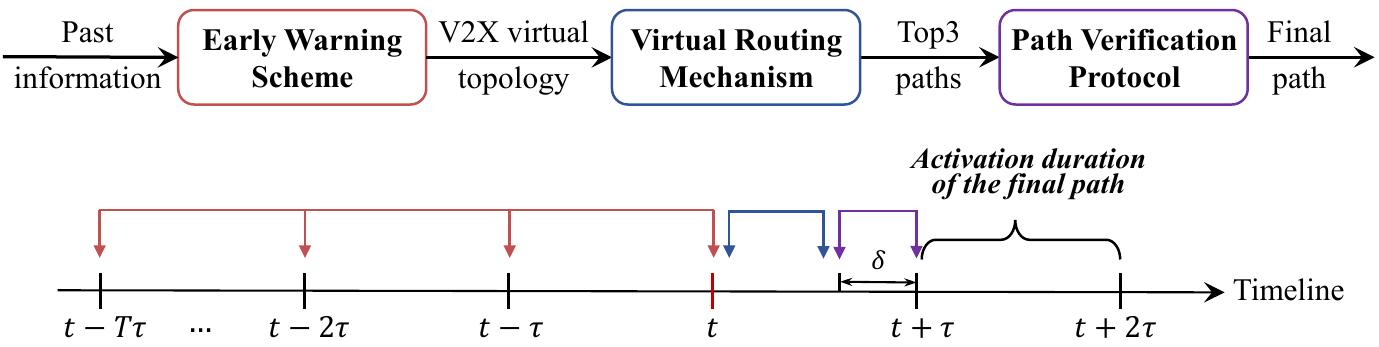}
	\caption{Overall framework of ROPE.}
	\vspace{-0.3cm}
	\label{fig-timeaxis}
\end{figure}

\section{Overall Framework of ROPE}	\label{sec-framework}
A multi-cell V2X communication scenario as exhibited in Fig.~\ref{fig-scene} is considered.
Base stations~(BSs) are connected to a cloud server through wired backhaul links. 
If a VUE is within the coverage range of more than one BSs, the VUE is associated with one BS from which the VUE receives the highest signal strength. Each VUE communicates with its associated BS via uplink~(direct V2I link), and it is able to communicate with other VUEs within the communication range via sidelink~(direct V2V link).
The cloud server keeps a database of historical reports on V2X communications. In the database, each item contains link type, transmitter information, receiver information, link strength, and traffic density. Link type is either V2I or V2V. Transmitter information is mobility and parameters of the transmitting VUE. Receiver information is either parameters of the associated BS or mobility and parameters of the receiving VUE. 
Link strength is the received signal strength of a V2X communication link. 
Traffic density is not obtained from the reports but retrieved by the cloud server and attached to each item.
From real-time surveillance of the intelligent transportation system~(ITS), the cloud server acquires recent traffic density, divides it into three levels~(low, medium, and high), and attaches it to the item.
Based on big data analysis, a link strength prediction model can be built for each type of V2X link on the cloud, which will be discussed later in Section \ref{sec-warning}.

The overall framework of ROPE is illustrated in Fig.~\ref{fig-timeaxis}. By default, a VUE directly communicates with a BS for obtaining the cloud service, and the cloud server collects VUE's mobility and parameters via the BS at a period of $\tau$.
At current time $t$, the cloud server predicts mobility of each vehicle and infers signal strength of each V2I link at $t+\tau$ based on the past information collected at time $t-i\tau$, where $i=0,1,...,T$. If the inferred strength does not exceed the RSS threshold or the predicted location is outside the coverage range of the BS, early warning is triggered for the corresponding VUEs and V2X virtual topology is built based on the inference results. On the basis of the virtual topology, the virtual routing mechanism is performed to seek top3 paths from warned VUEs to the cloud that satisfy QoS requirements. Shortly before activating the indirect paths, path verification protocol is executed at time $t+\tau-\delta$~($0 <\delta<\frac{\tau}{2}$) to assure that the final path is available and qualified. At time $t+\tau$, the final path is activated and path switchover is executed, which lasts until $t+2\tau$ and goes into the next cycle of the proposed ROPE method.

\section{Early Warning Scheme}	\label{sec-warning} 
In this section, a novel early warning scheme is proposed to proactively detect V2I link deterioration. 
First, a context-aware probabilistic neural network~(CAPNet) is designed and trained offline to learn probability distribution of V2I link strength. Next, VUE mobility feature and traffic density feature are fused to infer V2I link strength in an online fashion. Once the inferred strength is not above the RSS threshold or the predicted location is outside the coverage range of BSs, early warning is triggered and V2X virtual topology is formed based on the inferred link strength and predicted vehicle location.

\subsection{CAPNet Structure and Offline Training}
V2I link strength is mainly impacted by contexts of the communication pair, where impact factors include explicit factors~(2D location, antenna height, and velocity of VUE) and implicit factors~(traffic density around VUE). 
The former reveals 3D distance between VUE and BS, so it determines large-scale strength level.
The latter implies the number of vehicles around VUE.
Since multipath effect varies in different traffic density, the latter influences small-scale strength variation. 
To take both impact into account, a context-aware probabilistic neural network~(CAPNet) is deigned as the link strength prediction model.

\begin{figure}[t]
	\centering
	\includegraphics[width=0.8\columnwidth]{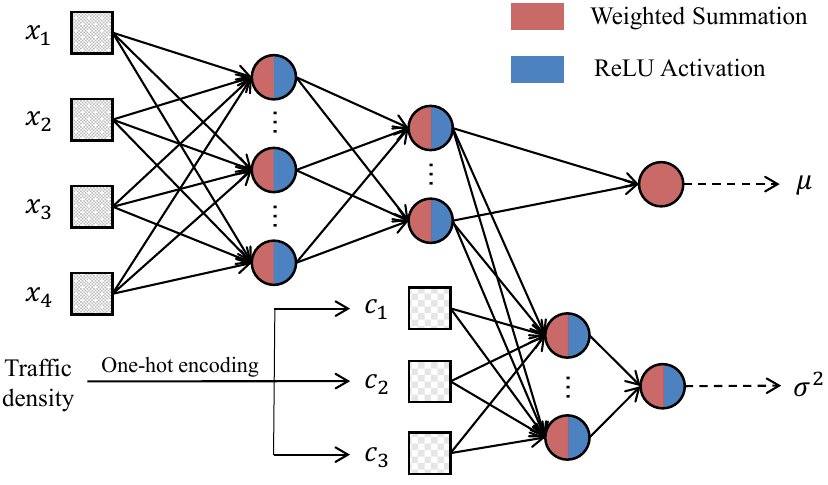}
	\caption{Structure of CAPNet.}
	\vspace{-0.3cm}
	\label{fig-net}
\end{figure}

As shown in Fig.~\ref{fig-net}, 2D location, antenna height, and velocity of a transmitting VUE are extracted from items in the database as explicit features $\mathbi{x}\in \mathbb{R}^4$. The traffic density after one-hot encoding acts as implicit features $\mathbi{c}\in \mathbb{R}^3$. 
To model probabilistic nature of RSS, CAPNet comprises two flows of fully-connected layers. One flow takes explicit features $\mathbi{x}$ as input and outputs RSS mean value $\mu$, which characterizes large-scale strength level. Another flow takes implicit features $\mathbi{c}$ as input and outputs RSS variance value $\sigma^2$, which characterizes small-scale strength variation. 
Since explicit features also imply whether a link is line of sight (LOS) or non-line of sight (NLOS) and influence small-scale signal strength, the flow of explicit features also joins the flow of implicit features to generate RSS variance value.
As such, probabilistic distribution of RSS is modeled by $p(y;\mu,\sigma^2)$, where $\mu,\sigma^2 = f_\theta(\mathbi{x},\mathbi{c})$ and $\theta$ denotes trainable parameters of CAPNet. To match the ground truth $\tilde{y}$ with this probability distribution, the negative log-likelihood~(NLL) loss function is to be minimized:
\begin{align}
 	\mathcal{L}(\theta) &= -\sum_{i=1}^{I}\log p(\tilde{y}_i;\mu_i,\sigma^2_i) \notag  \\
 	&= -\sum_{i=1}^{I}\log p(\tilde{y}_i;f_\theta(\mathbi{x}_i,\mathbi{c}_i)),
\end{align}
where $i$ denotes the $i$-th instance and $I$ is the number of instances in the training set. After optimization is finished, $\theta^\star$ is obtained and CAPNet is finalized on the cloud. Similarly, another CAPNet is built and optimized for V2V link, which is different in that 2D location, antenna height, and velocity of two communicating vehicles are extracted from the database as explicit factors $\mathbi{x}$ for modeling the V2V link strength.

\subsection{Feature Fusion and Online Inference}
At current time $t$, location and velocity of each VUE at time $t+\tau$ need to be foreseen for inferring V2I link strength at time $t+\tau$. Thereby, the cloud server executes mobility prediction for time $t+\tau$ based on the collected historical information (location and velocity at time $t-i\tau$, where $i=0,1,...,T$).
As the deep learning model MODA \cite{moda} captures dynamics-aware interaction between vehicles and facilitates multimodal future motions with definite goals, it is leveraged for accurate mobility prediction.
Since traffic density can be retrieved by the cloud and antenna height of VUEs is registered earlier, feature fusion is performed by combining such available information and the mobility prediction results to obtain explicit features $\mathbi{x}$ and implicit features $\mathbi{c}$. When CAPNet is applied to these fused features, V2I link strength at $t+\tau$ can be evaluated by $\mu,\sigma^2 = f_{\theta^\star}(\mathbi{x},\mathbi{c})$ proactively. 
Under the RSS threshold $\gamma_{th}$, if $\mu-\sigma \leq \gamma_{th}$ holds for a VUE, V2I link deterioration is detected and early warning is triggered for this VUE, which is called a warned VUE. Although the mean value represents average link status, the standard deviation is subtracted to find more possible warned VUEs and improve V2I link quality of these VUEs.
Additionally, if the predicted location of a VUE falls out of coverage of its associated BS, early warning is also triggered and this VUE is also a warned VUE. 
Otherwise, the cloud server waits until the next time stamp $t+\tau$ and continues to predict vehicle mobility and infer V2I link strength at time $t+2\tau$.

\subsection{V2X Virtual Topology Formation}
If there exists any warned VUE in the system, V2X virtual topology is formed on the basis of the inference results, which include predicted location of each VUE and inferred mean strength of each V2I link at time $t+\tau$.
Specifically, V2I link is virtually existent if the inferred mean strength is greater than the RSS threshold $\gamma_{th}$ and the VUE is in the coverage range $d_I$ of its associated BS. As predicted location of all VUEs is known, potential V2V pairs are identified if their predicted distance is less than the communication range $d_V$. Next, feature fusion and online inference are carried out for these V2V pairs by the V2V-customized CAPNet, which produces mean strength of V2V links at time $t+\tau$.
V2V link is virtually existent if the inferred mean strength is greater than the RSS threshold $\gamma_{th}$.
After combining virtually existent V2I and V2V links, V2X virtual topology $\mathcal{G}=(\mathcal{V},\mathcal{E})$ is constructed, where $\mathcal{V}$ is the node set containing all VUE nodes and one node representing all BSs, and $\mathcal{E}$ is the edge set accommodating all virtual V2X links.

\section{Virtual Routing Mechanism}	\label{sec-routing}
This section describes the routing mechanism applied to the V2X virtual topology. First, QoS-related link and path metrics in terms of strength, connectivity, and hop count are designed. Second, given these metrics, a restricted widest path problem is formulated and proven to be NP-complete. Third, a top3 routing algorithm~(TORA) is developed to solve the NP-complete problem under acceptable computational complexity and yield three best feasible paths.
\subsection{QoS Metric Design}
\subsubsection{Link normalized strength}
it is defined as the inferred mean strength normalized by RSS threshold $\gamma_{th}$ and RSS maximum $\gamma_M$. 
Given the inferred mean strength $\mu$ of the corresponding V2X link at time $t+\tau$, the link normalized strength $l_S$ is calculated as
\begin{equation} \label{eq-rate}
	l_S = \frac{\mu-\gamma_{th}}{\gamma_M-\gamma_{th}},
\end{equation}
where $l_S \in (0,1]$ and the larger the better. 

\begin{figure}[t]
	\centering
	\subfigure[$\alpha \in \text{\big[}0, \frac{\pi}{2}\text{\big]}$ ] { 
		\label{subfig-dur1}
		\includegraphics[width=0.32\columnwidth]{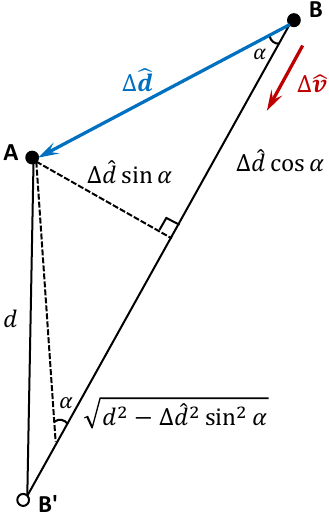}}
	\vspace{0.3cm}
	\subfigure[$\alpha \in \big(\frac{\pi}{2}, \pi \text{\big]}$] { 
		\label{subfig-dur2}
		\includegraphics[width=0.64\columnwidth]{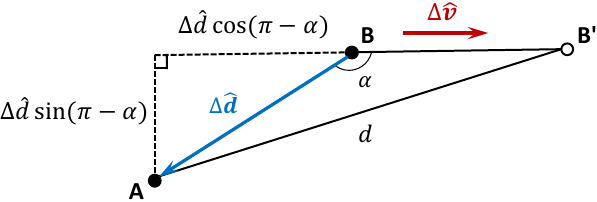}}
	\caption{Calculation of link duration time.}
	\label{fig-duration}
\end{figure}

\subsubsection{Link connectivity} 
it is defined as the ratio of duration time when a direct communication link between two nodes will stay continuously connected or available from $t+\tau$ to $t+2\tau$.

The geometric illustration of link duration time calculation is shown in Fig.~\ref{fig-duration}. 
Let $\Delta \hat{\mathbi{d}}$ denote the predicted relative displacement between node A and node B at time $t+\tau$ (the scalar form is $\Delta\hat{d}$), $\Delta\hat{\mathbi{v}}$ denote the predicted relative velocity of the two nodes at time $t+\tau$ (the scalar form is $\Delta \hat{v}$), and $\alpha \in [0,\pi]$ indicate the angle between $\Delta \hat{\mathbi{d}}$ and $\Delta\hat{\mathbi{v}}$.
For the link between node $\text{A}$ and node $\text{B}$ at time $t+\tau$, assume that $\text{A}$ is relatively static to $\text{B}$, and $\text{B}^\prime $ is the critical position where node $\text{B}$ is about to travel out of the communication range $d$ of node $\text{A}$. Here, $d$ is either $d_I$ for V2I links or $d_V$ for V2V links. Hence, the travel time from $\text{B}$ to $ \text{B}^\prime $ indicates the link duration time between node $\text{A}$ and node $\text{B}$. 
For $\alpha \in \text{\big[}0, \frac{\pi}{2}\text{\big]}$, the link duration time can be computed as 
\begin{equation} \label{eq:dur1}
	\hat{T} = \frac{\sqrt{d^2 - \Delta{\hat{d}}^2 \sin^2 \alpha} + \Delta\hat{d} \cos \alpha }{\Delta \hat{v}}.
\end{equation}
For $\alpha \in \big(\frac{\pi}{2}, \pi \text{\big]}$, the link duration time can be computed as 
\begin{align} \label{eq:dur2}
	\hat{T} &= \frac{\sqrt{d^2 - \Delta{\hat{d}}^2 \sin^2(\pi-\alpha)} - \Delta\hat{d} \cos(\pi-\alpha) }{\Delta \hat{v}} \notag \\
	&=  \frac{\sqrt{d^2 - \Delta{\hat{d}}^2 \sin^2 \alpha} + \Delta\hat{d} \cos \alpha }{\Delta \hat{v}},
\end{align}
which is the same as Eq.~(\ref{eq:dur1}). Given the link duration time $\hat{T}$, link connectivity from $t+\tau$ to $t+2\tau$ is determined after normalization:
\begin{equation}	\label{eq:connect}
l_C=
\begin{cases}
	\hat{T}/\tau, \qquad & \mathrm{if} \quad \hat{T}<\tau, \\
	1, \qquad & \text{otherwise}, \\
\end{cases}
\end{equation}
where $l_C \in (0,1]$ and the larger the better. 

\subsubsection{Link hop count}
it is set to 1 for each virtual edge that represents either a V2I link or a V2V link in the V2X virtual topology, i.e., $l_H = 1$.

Given the above link metrics, path metrics can be obtained. Assume that a path $J$ from node $i$ to node $m$ contains the link set $\{(i,j),(j,k),..., (l,m)\}$. The link normalized strength, link connectivity, and link hop count between node $i$ and node $j$ are represented by $l_S(i,j)$, $l_C(i,j)$, and $l_H(i,j)$, respectively. 
Similarly to~\cite{wang1996quality}, path strength $p_S(J)$ and path connectivity $p_C(J)$ can be calculated by the concave rule, while path hop count $p_H(J)$ can be computed by the additive rule:
\begin{align}
	p_S(J) & =\min \left\{l_S( i,j ) ,l_S( j,k ) ,\cdots ,l_S( l,m ) \right)\}, \\
	p_C(J) & =\min \left\{l_C( i,j ) ,l_C( j,k ) ,\cdots ,l_C( l,m ) \right)\}, \\
	p_H(J) & =\sum \left\{l_H( i,j ) ,l_H( j,k ) ,\cdots ,l_H( l,m ) \right)\},
\end{align}
where $p_S(J) \in (0,1]$, $p_C(J) \in (0,1]$, and $p_H(J) \in (0,+\infty)$.

\subsection{Problem Formulation}	
For acquiring multimedia service with QoS provision, the goal is to find three paths from a warned VUE to a BS on the V2X virtual topology that maximize the path strength while satisfying the connectivity and the hop count constraints. Therefore, three problems are formulated, which aim to find the best feasible path $J_1$, the second best feasible path $J_2$, and the third best feasible path $J_3$ from the path set $\mathcal{J}$. Here, \textit{best} implies that the highest path strength is achieved and 
\textit{feasible} means that both the connectivity constraint $C_{th}$ and the hop count constraint $H_{th}$ are satisfied. The path set $\mathcal{J}$ includes all the possible paths from the source node $s$ (the warned VUE) to the destination node $d$ (the BS) on the topology graph $\mathcal{G}$.

\textit{Problem 1:}
\begin{align}
	& J_1 = \text{argmax}_{J \in \mathcal{J}}~p_S(J), \\
	\text{s.t.} \qquad &p_C(J) > C_{th}, \label{eq:rconstr} \\
	\quad &p_H(J) < H_{th}. \label{eq:dconstr}
\end{align}

\textit{Problem 2:}
\begin{align}
	& J_2 = \text{argmax}_{J \in \mathcal{J}-\{J_1\} }~p_S(J), \\
	\text{s.t.}\qquad &\text{Eq}.(\ref{eq:rconstr})-(\ref{eq:dconstr}) \notag.
\end{align}

\textit{Problem 3:}
\begin{align}
	& J_3 = \text{argmax}_{J \in \mathcal{J}-\{J_1, J_2\} }~p_S(J), \\
	\text{s.t.}\qquad &\text{Eq}.(\ref{eq:rconstr})-(\ref{eq:dconstr}) \notag.
\end{align}

\begin{theorem}
	Problem 1 is a restricted widest path problem, which is equivalent to a restricted shortest path problem and is NP-complete.
\end{theorem}

\begin{IEEEproof}
The path set $\mathcal{J}$ is based on the V2X virtual topology graph $\mathcal{G}=(\mathcal{V},\mathcal{E})$. After pruning the links $(i, j) \in \mathcal{E}$ that has $l_C(i, j) \leq C_{th}$ and removing the disconnected nodes, the original graph $\mathcal{G}=(\mathcal{V},\mathcal{E})$ is trimmed to a new graph
$\mathcal{G}^\prime=(\mathcal{V}^\prime,\mathcal{E}^\prime)$.  Since $l_C(i,j) > C_{th}$ holds for any link $ (i,j) \in \mathcal{E}^\prime$, given $p_C(J) =\min \left\{l_C(i,j) ,l_C(j,k) ,\cdots ,l_C(l,m)\right\}$, the new path set $\mathcal{J}^\prime$ on the new graph $\mathcal{G}^\prime$ satisfies that $p_C(J) > C_{th}$ for all $J \in \mathcal{J}^\prime$ . Consequently, the connectivity constraint can be eliminated and Problem 1 is converted into

\textit{Problem 4:}
\begin{align}
	& J_1 = \text{argmax}_{J \in \mathcal{J}^\prime}~p_S(J), \\
	\text{s.t.} \qquad &\text{Eq}.(\ref{eq:dconstr}), \notag
\end{align}
where $p_S(J) =\min \left\{l_S( i,j ) ,l_S( j,k ) ,\cdots ,l_S( l,m ) \right)\}$ further expands $J_1 = \text{argmax}_{J \in \mathcal{J}^\prime}\text{min}_{(i,j)\in J} l_S(i,j)$. Regarding $l_S(i,j)$ as the width of the link $(i,j)$ and the minimum link width along a path $J$ as the width of the path, the above problem can be interpreted as finding the widest path under the hop count constraint. Thereby, it is defined as the restricted widest path problem. Since $l_S(i,j) \in (0,1]$ yields $g_S(i,j) = \frac{1}{l_S(i,j)} \in [1,+\infty)$, the objective $J_1 = \text{argmax}_{J \in \mathcal{J}^\prime}\text{min}_{(i,j)\in J} l_S(i,j)$ is equivalent to $J_1 = \text{argmin}_{J \in \mathcal{J}^\prime}\text{max}_{(i,j)\in J} g_S(i,j)$. By utilizing the property of $L_\infty$ norm, the latter part $\text{max}_{(i,j)\in J} g_S(i,j)$ equals $\left(\sum_{(i,j)\in J} g_S(i,j)^{\lambda} \right)^{1/\lambda}$ as $\lambda \rightarrow +\infty$. Hence, $J_1 = \text{argmin}_{J \in \mathcal{J}^\prime} \left(\sum_{(i,j)\in J} g_S(i,j)^{\lambda} \right)^{1/\lambda} = \text{argmin}_{J \in \mathcal{J}^\prime} \sum_{(i,j)\in J} g_S(i,j)^{\lambda}$. Let $h_S(i,j) = g_S(i,j)^\lambda$, and the restricted widest path problem is equivalent to the problem:
\begin{align}
	& J_1 = \text{argmin}_{J \in \mathcal{J}^\prime} \sum_{(i,j)\in J} h_S(i,j), \\
	\text{s.t.} \qquad &\text{Eq}.(\ref{eq:dconstr}). \notag
\end{align}
When regarding $h_S(i,j)$ as the length of the link $(i,j)$ and the sum of the link length along a path $J$ as the length of the path, the problem above is identical to the restricted shortest path problem, which is proved to be NP-complete~\cite{gary1979computers}. Therefore, Problem 1 is equivalent to the restricted shortest path problem, which is also NP-complete.
\end{IEEEproof}

Similarly, finding paths from the path set $\mathcal{J}^\prime$ on the new graph $\mathcal{G}^\prime$, Problem 2 and 3 can be transformed into Problem 5 and 6, which are also NP-complete.

\textit{Problem 5:}
\begin{align}
	& J_2 = \text{argmax}_{J \in \mathcal{J}^\prime-\{J_1\}}~p_S(J), \\
	\text{s.t.} \qquad &\text{Eq}.(\ref{eq:dconstr}). \notag
\end{align}

\textit{Problem 6:}
\begin{align}
	& J_3 = \text{argmax}_{J \in \mathcal{J}^\prime-\{J_1, J_2\}}~p_S(J), \\
	\text{s.t.} \qquad &\text{Eq}.(\ref{eq:dconstr}). \notag
\end{align}

As such, Problem 4-6 need to be solved to find three best feasible paths.

\subsection{Top3 Path Routing Algorithm}
To solve the NP-complete problems under acceptable computational complexity, a top3 path routing algorithm (TORA) is developed to find three best feasible paths as formulated in Problem 4-6. Generally, TORA is composed of two main procedures: widest feasible path finding~(WFPF) and deviation path ranking~(DPF). The WFPF procedure aims to find the widest feasible path for a given communication pair (solve Problem 4), while the DPR procedure deletes the obtained widest feasible path from the search space and tries to find the next widest feasible path for the communication pair (solve Problem 5 and 6 iteratively). Details of the two procedures are elaborated below.

\begin{algorithm}[t]
	\caption{Widest Feasible Path Finding}
	\label{alg:wfpf}
	\begin{algorithmic}[1]	
		\State Backward\_Dijkstra($\mathcal{G}^\prime$, $d$);	\Comment{For feasibility check}
		\If {$b[s]\geq H_{th}$}
		\State Return failure	
		\EndIf
		\State Forward\_Dijkstra($\mathcal{G}^\prime$, $s$, $H_{th}$);
		\Comment{For optimality search}
		\If {$f[d] < H_{th}$}
		\State Return success
		\Else
		\State Return failure
		\EndIf		
	\end{algorithmic}
\end{algorithm}

\begin{algorithm}[t]
	\caption{Relaxation Procedure in Forward\_Dijkstra}
	\label{alg:relax}
	\begin{algorithmic}[1]	
		\Procedure{Relax}{$u, v$}
		\State Let $tmp$ be a temporary node;
		\State $w[tmp]=\min(w[u],l_S(u,v))$;
		\State $f[tmp]= f[u] + l_H(u,v)$;
		\State $b[tmp]= b(v)$;
		\If{PREFER$(tmp, v) == tmp$}
		\State $w[v]=w[tmp]$;
		\State $f[v]=f[tmp]$;		
		\State $\pi[v]=u$;
		\EndIf		
		\EndProcedure
	\end{algorithmic}
\end{algorithm}

\begin{algorithm}[t]
	\caption{Preference Rule in Forward\_Dijkstra}
	\label{alg:prefer}
	\begin{algorithmic}[1]	
		\Procedure{Prefer}{$x, y$}
		\If{$w[x] > w[y]~\text{and}~f[x] +b[x] < H_{th}$}
		\State Return $x$
		\ElsIf{$w[x] < w[y]~\text{and}~f[y] +b[y] < H_{th}$}
		\State Return $y$
		\ElsIf{$f[x] +b[x] < f[y] +b[y]$}
		\State Return $x$
		\EndIf
		\State Return $y$
		\EndProcedure
	\end{algorithmic}
\end{algorithm}

\subsubsection{Widest Feasible Path Finding~(WFPF)} 
It is inspired by a two-directional heuristic algorithm~\cite{korkmaz2001multi} in multi-constrained optimal path selection, 
which is tailored to fit our restricted widest path problem. The pseudocode of WFPF is shown in Algorithm~\ref{alg:wfpf} and explained as follows. 
For feasibility check, WFPF calculates the least hop count that each node spends to reach the destination node. To achieve this, it takes the graph $\mathcal{G}^\prime$ and the destination node $d$ as input, with link hop count $l_H$ resembling the link length. The shortest path algorithm Dijkstra is run to find the least hop count path from node $d$ to each node, which is wrapped into $\text{Backward\_Dijkstra}(\mathcal{G}^\prime, d)$. After executing $\text{Backward\_Dijkstra}$, each node $u$ is associated with an item $b[u]$ recording the least hop count that node $d$ spends to reach it and vice versa. 
If the source node has $b[s] \geq H_{th}$, no feasible paths can be found for the communication pair $(s, d)$ since even the lower bound of path hop count exceeds the hop count constraint. 
For optimality search, WFPF explores paths from the source node $s$ with the target of maximizing the path strength while satisfying the hop count constraint. In addition to $b[u]$, each node $u$ maintains three more items: $w[u], \pi[u],~\text{and}~f[u]$. $w[u]$ represents the path strength, resembling the path width of the optimal path from node $s$ to node $u$. $\pi[u]$ records the predecessor of $u$ on the optimal path. $f[u]$ denotes the path hop count of the optimal path. Another Dijkstra algorithm is run in a forward direction, which takes the graph $\mathcal{G}^\prime$, the source node $s$, and the hop count constraint $H_{th}$ as input. Hence, it is wrapped into $\text{Forward\_Dijkstra}(\mathcal{G}^\prime, s, H_{th})$.  $\text{Forward\_Dijkstra}$ initially sets $w[u]=0, \pi[u]=NIL, f[u] = 0~\text{for}~u \neq s$, and sets $w[s]=+\infty, \pi[s]=NIL, f[s] = 0$. Starting from the source node $s$, the node with maximum $w[\cdot]$ (e.g., $u$) is recurrently extracted from the unvisited node set and the relaxation procedure is applied to the neighbor node $v$ of $u$, which is shown in Algorithm~\ref{alg:relax}. Whether $v$ takes $u$ as its predecessor and updates its items depend on the preference rule in Algorithm~\ref{alg:prefer}. 
Notably, $f[v]$ is current path hop count from the source node $s$ to $v$, and $b[v]$ is the least path hop count from $v$ to the destination node $d$. $f[v] + b[v] < H_{th}$ is necessary for $v$ to be on a feasible path from $s$ to $d$, so the backward result helps the forward search foresee the feasibility of nodes on the path.
After all nodes of $\mathcal{G}^\prime$ are visited, $\text{Forward\_Dijkstra}$ is finished and items of each node are fixed. If the destination node has $f[d] < H_{th}$, the best feasible path $J_1$ is found from $s$ to $d$ successfully. Otherwise, no feasible path is found. As a result, Problem 4 is solved by WFPF.

\begin{figure*}[t]
	\centering
	\includegraphics[width=15cm]{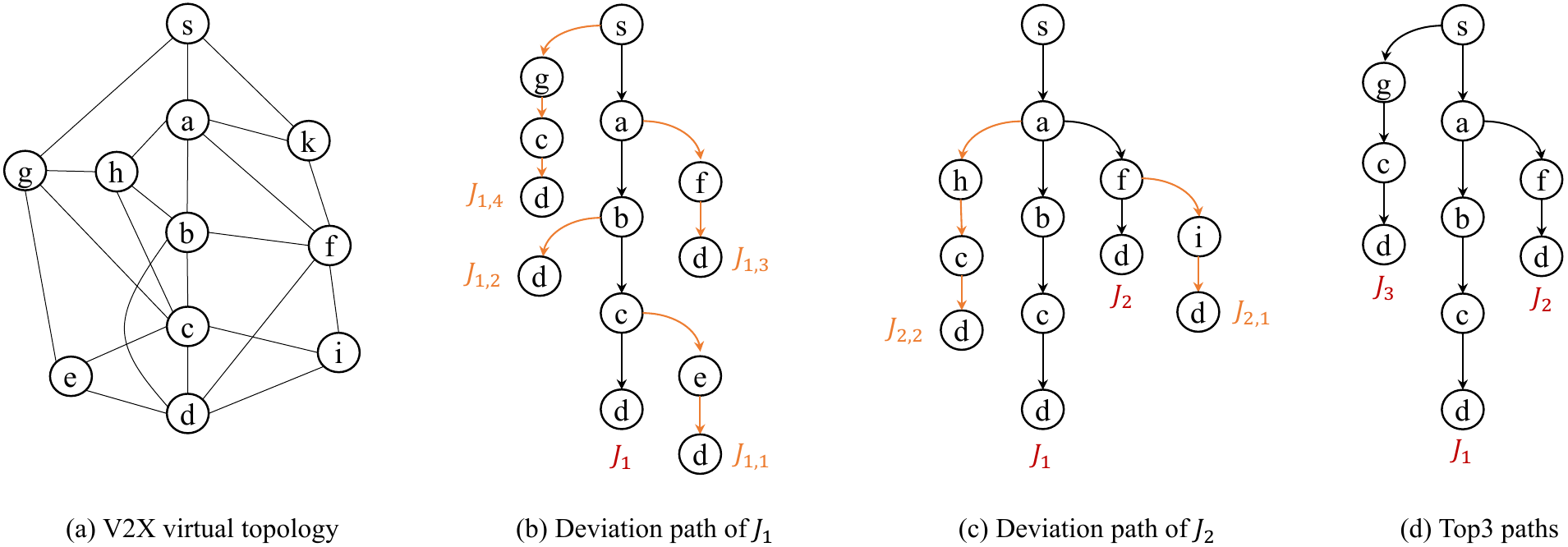}
	\caption{Example of deviation path ranking. The path depicted in orange is the best feasible deviation path found by Forward\_Dijkstra, and the symbol highlighted in red is the best feasible complete path provided by WFPF or generated by ranking the candidate set.}
	\vspace{-0.3cm}
	\label{fig-deviation}
\end{figure*}
	
\subsubsection{Deviation Path Ranking~(DPR)}
Given the best feasible path $J_1$, deviation path ranking is conducted to find the second best feasible path $J_2$ and the third best feasible path $J_3$. 
In DPR, all nodes on $J_1$ except the destination node $d$ are deemed as branch nodes of $J_1$. For each branch node $u$, the root path is the path segment from $s$ to $u$ on $J_1$ (the root path of $s$ is none). The deviation paths are the paths from $u$ to $d$ without other nodes on its root path and the outgoing link of $u$ on $J_1$. The root path and each deviation path make up a complete path from $s$ to $d$ via $u$, and the complete paths of all branch nodes exactly form the path set $\mathcal{J}^\prime - \{J_1\}$. 
Searching the best feasible path from $\mathcal{J}^\prime - \{J_1\}$ for obtaining $J_2$ is decomposed into finding the best feasible deviation path for each branch node, ranking the corresponding complete paths with respect to the path strength, and outputting the complete path with the highest path strength as $J_2$.
To find the best feasible deviation path of each branch node, $\text{Forward\_Dijkstra}$ is reused with modified input. 
For branch node $u$, a masked graph $G^\prime_{u,J_1}$ is obtained after masking other nodes on its root path and the outgoing link of $u$ on $J_1$ from $\mathcal{G}^\prime$. $D_{u,J_1}$ is calculated by subtracting $H_{th}$ from the hop count of $u$'s root path as the customized hop count constraint for the deviation path. In this way, running $\text{Forward\_Dijkstra}(G^\prime_{u,J_1}, u, D_{u,J_1})$ produces the best feasible deviation path of branch node $u$ to the destination node $d$. By concatenating the root path of $u$ with this deviation path, a complete path from $s$ to $d$ via $u$ is generated and added to a candidate set. After the best feasible deviation paths of all branch nodes are added to the set, rank the paths descendingly with respect to path strength and pop up the path with the highest path strength as the second best feasible path $J_2$. 
Similarly, to find the third best feasible path $J_3$, branch nodes of 
$J_2$ are identified and $\text{Forward\_Dijkstra}(G^\prime_{v,J_2}, v, D_{v,J_2})$ is applied to each branch node $v$ of $J_2$. Here, a masked graph $G^\prime_{v,J_2}$ is obtained after masking other nodes on its root path and the outgoing link of $v$ on $J_2$ from $\mathcal{G}^\prime$. Also, if the root path of $v$ is contained in $J_1$, the outgoing link of $v$ on $J_1$ is also masked. 
After executing Forward\_Dijkstra for each branch node, the best feasible path of each branch node is produced and added to the aforementioned candidate set. The paths in the set are ranked descendingly with respect to path strength and the path with the highest path strength is popped up as the third best feasible path $J_3$. An example of DPR is shown in Fig.~\ref{fig-deviation}. As a result, Problem 5-6 is solved by DPR. 

In view of the procedures above, complexity of TORA is analyzed as follows.
Let $N$ be the number of nodes and $M$ be the number of edges in $\mathcal{G}^\prime$. Given that complexity of Dijkstra algorithm with Fibonacci heap implementation~\cite{fredman1987fibonacci} is $O(N\log N + M)$, both $\text{Backward\_Dijkstra}$ and $\text{Forward\_Dijkstra}$ in WFPF have complexity of $O(N\log N + M)$. In DPR, if there are $l_1$ branch nodes on $J_1$, it calls $l_1$ times $\text{Forward\_Dijkstra}$ to find $J_2$ and hence has complexity of $O\left(l_1 (N\log N + M)\right)$. If there are $l_2$ branch nodes on $J_2$, it calls $l_2$ times $\text{Forward\_Dijkstra}$ to find $J_3$ and hence has complexity of $O\left(l_2 (N\log N + M)\right)$. Therefore, the overall complexity of TORA is $O\left((l_1+l_2+2) (N\log N + M)\right)$. Since the worst-case value of $l_1$ and $l_2$ in a condensed graph is $N$. The worst-case complexity of TORA is $O\left(N (N\log N + M)\right)$, which is acceptable in solving three NP-complete problems.

\section{Path Verification Protocol} \label{sec-verification}
After virtual routing mechanism is applied to the V2X virtual topology, top3 paths are determined and one of these paths is selected for activation at time $t+\tau$. Since routing decisions are made based on the predicted link status, some of top3 paths may be unavailable or unqualified due to imperfect prediction. Thereby, before switching from a direct V2I path to an indirect multi-hop path, availability and quality of top3 paths need to be verified and the best path in real-world situation should be activated. 
Nevertheless, high mobility of vehicles makes actual topology vary rapidly. How to verify top3 paths shortly before switchover and react to the situation when all top3 paths are ineligible are challenging. 
To address these issues, a path verification protocol is designed with three key operations: link check, path check, and path mending. 

\textbf{Link Check.}
When a path $J_k$ of the top3 paths needs verification ($k=1,2,3$), the cloud server informs the nodes on this path of path ID, pre-hop node ID, and post-hop node ID at time $t+\tau-\delta_k$, and then starts a timer.
For nodes on the paths that are out of coverage of BSs, the in-coverage nodes on the paths help convey the control information between the out-of-coverage nodes and the cloud server.
The nodes transmit link check message (LCM) to its post-hop node simultaneously, which contains information about their current location and velocity. Once a node receives LCM from its pre-hop node, it measures signal strength and calculates link duration time. 
The calculation of link duration time is similar to Eq.~(\ref{eq:dur1}) and Eq.~(\ref{eq:dur2}) except that relative location and velocity of two nodes are based on real information at time $t+\tau-\delta_k$ rather than predicted mobility at $t+\tau$. 
Consequently, the duration time minus $\delta_k$ indicates the estimated link duration time from $t+\tau$, which is normalized according to Eq.~(\ref{eq:connect}) for computing link connectivity.
If link strength is above the RSS threshold $\gamma_{th}$ and link connectivity is greater than the connectivity constraint $C_{th}$, the link is deemed qualified and the node reports link strength and connectivity to the cloud server. Otherwise, the link is deemed unqualified and no report is sent. 

\textbf{Path Check.}
After the timer expires, the cloud server identifies qualified and unqualified links of $J_k$ based on the collected reports and adds the unqualified links to a fault set. If all links are qualified, then path strength, path connectivity,  and path hop count are sure to satisfy the QoS requirements. Hence, the path is deemed qualified and chosen as the final path at $t+\tau$, which terminates the path verification protocol. Otherwise, the path with any unqualified link is deemed as an unqualified path.
When path $J_k$ is unqualified, the cloud server turns to the next path $J_{k+1}$. If $J_{k+1}$ has links that have been added to the fault set, the path is deemed unqualified without the link check operation, and the cloud server turns to the next path $J_{k+2}$. Since link check is only performed for the subsequent paths without links in the fault set, signaling consumption can be saved and verification process can be accelerated. If $J_{k+1}$ does not have links in the fault set, the nodes on $J_{k+1}$ are informed at $t+\tau-\delta_{k+1}$ and link check is conducted in a similar way, and so is path check.

\begin{figure}[t]
	\centering
	\subfigure[Successful path mending] { 
		\label{fig-mend1}
		\includegraphics[width=0.8\columnwidth]{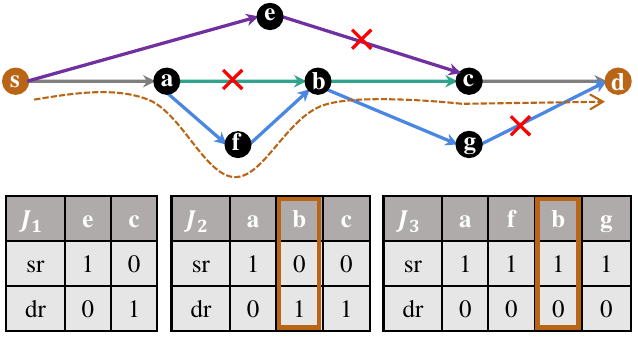}}
	\vspace{0.2cm}
	\subfigure[Unsuccessful path mending] { 
		\label{fig-mend2}
		\includegraphics[width=0.8\columnwidth]{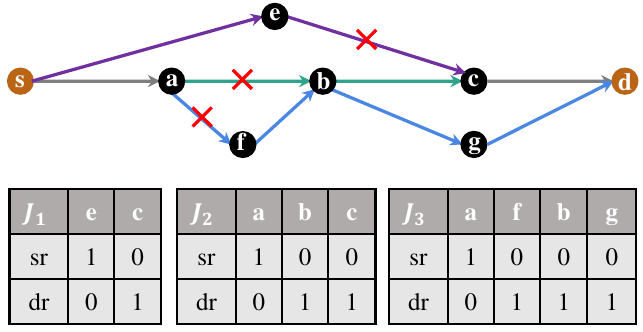}}
	\caption{Examples of path mending.}
	\label{fig-mend}
\end{figure}

\textbf{Path Mending.}
When none of top3 paths are deemed qualified, a backup plan is launched on the cloud. If link check is performed for more than one path, a path mending procedure is carried out quickly to generate a qualified path out of two unqualified paths. Otherwise, all top3 paths are verified to be unsuccessful and the direct V2I path is still utilized at $t+\tau$ if it is available. The key idea of path mending is elucidated below. For a path $J_k$ after link check, the cloud server annotates each node $v$ on the path except the source node $s$ and the destination node $d$ with two flags $\text{sr}_k[v]$ and $\text{dr}_k[v]$. $\text{sr}_k[v]$ is 1 if the source node is reachable from $v$. Otherwise, it is set to 0. $\text{dr}_k[v]$ is 1 if the destination node is reachable from $v$. Otherwise, it is set to 0. 
Here, \textit{reachable} means that the subpath of $J_k$ from $s$ to $v$ or $v$ to $d$ satisfies the requirements of path strength and path connectivity.
When two unqualified paths $J_e$ and $J_f$ have shared nodes, there is possibility of a shared node $u$ having $\text{sr}_e[u]=1~\&~ \text{dr}_f[u]=1$ or $\text{sr}_f[u]=1~\&~\text{dr}_e[u]=1$. Then, two subpaths from the two unqualified paths can form an available path $J_{ef}$ or $J_{fe}$ from $s$ to $d$. The path hop count is calculated on the new path and feasibility is confirmed when it is less than the hop count constraint. All such new paths are generated after path mending, and the feasible path with the highest path strength is selected as the final path to be activated at time $t+\tau$. If no feasible paths are generated and the direct V2I path is available, the direct path is activated rather than the indirect multi-hop V2X path. 
For better illustration, two examples of path mending are shown in Fig.~\ref{fig-mend}, where the three paths between the source node $s$ and the destination node $d$ in each example are $J_1, J_2, J_3$ from top to bottom. 
In Fig.~\ref{fig-mend1}, the shared node $b$ on path $J_2$ and path $J_3$ satisfies $\text{sr}_3[b]=1~\&~\text{dr}_2[b]=1$, so the subpath from node $s$ to $b$ of path $J_3$ and the subpath from node $b$ to $d$ of path $J_2$ can be successfully mended. 
In Fig.~\ref{fig-mend1}, the shared node $b$ on path $J_2$ and path $J_3$ as well as the shared node $c$ on path $J_1$ and path $J_2$ cannot mend two unqualified paths into one qualified path, so the direct V2I path is activated.

\begin{figure}[t]
	\centering
	\includegraphics[width=7cm]{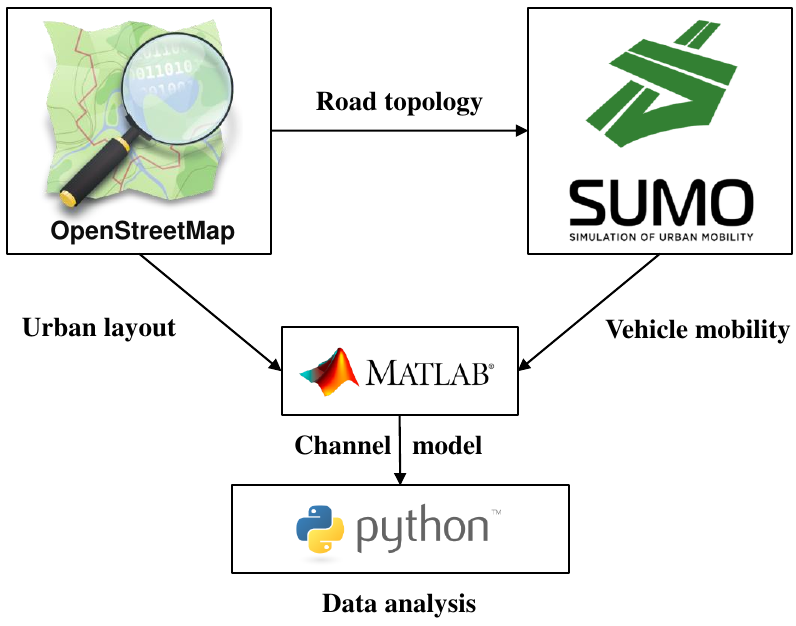}
	\caption{Simulation framework.}
	\label{fig-simulation}
\end{figure}

\section{Performance Evaluation} \label{sec-evaluation}
In this section, experiments are carried out to substantiate effectiveness of the proposed ROPE method. Particularly, the simulation framework integrating multiple tools is first described. The simulated urban scenario and configuration parameters are then introduced. Next, the early warning scheme is validated by analyzing the impact factors of V2I links and evaluating successful warning ratio of different prediction methods. Subsequently, robustness to link failures is validated by comparing the proposed method with and without remedy solutions. Finally, comparison with baseline methods is conducted by evaluating path strength and path qualification ratio.

\subsection{Simulation Framework}
To simulate realistic V2X communications, several tools are combined in the proposed framework, as shown in Fig.~\ref{fig-simulation}. Particularly, OpenStreerMap (OSM) \cite{OpenStreetMap} is an open geographic database, which provides road topology and urban layout (buildings and vegetation) in real-world cities. 
SUMO \cite{SUMO2018} is a microscopic and continuous traffic simulation package, which takes road topology from OSM as input and generates mobile traces of vehicles at different traffic density. 
$\text{GEMV}^2$ \cite{boban2014geometry,aygun2016geometry} is an efficient geometry-based channel model for V2X communications, which is implemented in MATLAB and utilizes outlines of vehicles, buildings, and vegetation to produce location-aware link-level modeling, hence striking a balance between accuracy and complexity.
Fed with urban layout from OSM and vehicle mobility from SUMO, $\text{GEMV}^2$ classifies each link into three groups: LOS, NLOS due to building/foliage (NLOSb), and NLOS due to vehicles (NLOSv), and the computation of link strength varies for different groups.
After V2X links and their signal strength are provided by MATLAB, Python is leveraged to conduct offline data analysis as well as online workflow of early warning scheme, virtual routing mechanism, and path verification protocol.

\subsection{Scenario and Parameters}
\begin{figure}[t]
	\centering
	\subfigure[New York Manhattan] { 
		\label{subfig:osm}
		\includegraphics[width=0.407\columnwidth]{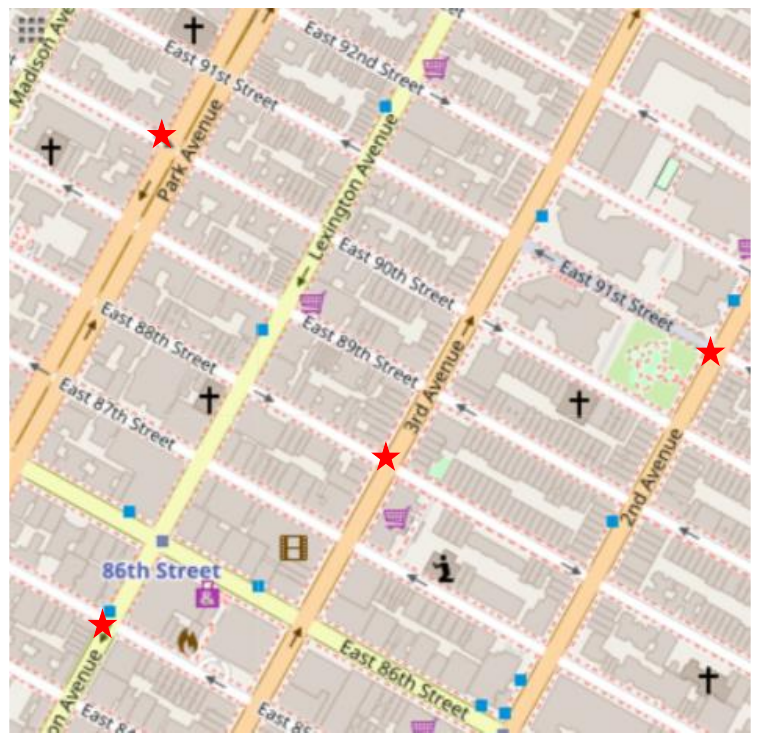}}
	\vspace{0.2cm}
	\subfigure[RSS heatmap] { 
		\label{subfig:heat}
		\includegraphics[width=0.48\columnwidth]{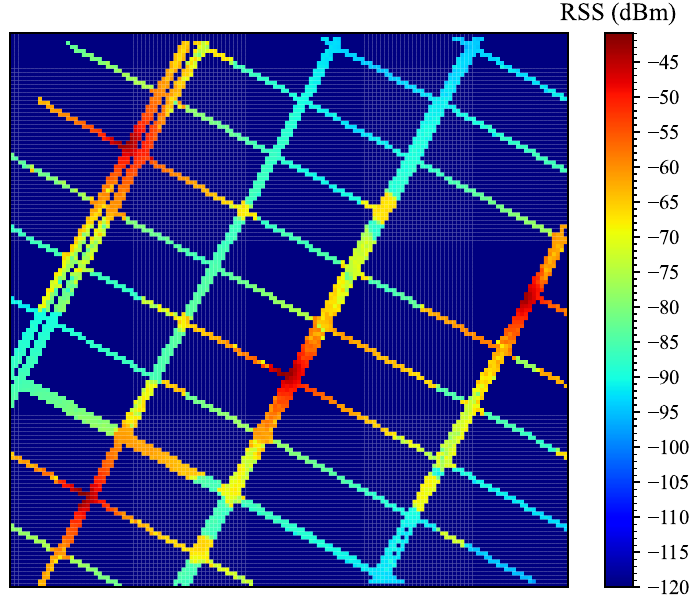}}
	\caption{Urban scenario.}
	\label{fig-city}
\end{figure}

The simulated urban scenario is illustrated in Fig.~\ref{subfig:osm}, where a 696 m $\times$ 700m area is selected from Manhattan, New York city. 
Four BSs are deployed in the area and depicted as red stars. 
Specifically, 4 GHz carrier frequency with 30 kHz subcarrier spacing is adopted. All BSs and VUEs are equipped with omni-directional antennas.
The configuration parameters of BSs comply with those of micro BSs in the urban grid scene \cite{37.885}, with fixed antenna height of 5 m and transmit power of 24 dBm. 
There are two types of VUEs: $85\%$ passenger vehicle and $15\%$ truck/bus. The former has antenna height of 1.6 m and the latter has antenna height of 3.1 m. The transmit power of VUEs is 23 dBm and the noise power is -114 dBm. 
The maximum RSS value is set to $\gamma_M=-10$ dBm.
The coverage range $d_I$ of BSs is 400 m and the communication range $d_V$ between VUEs is 300 m. If a vehicle is within the coverage range of multiple BSs, it is associated with the BS from which the VUE receives the highest signal strength.
The connectivity constraint $C_{th}$ is set to 0.999 and the hop count constraint $H_{th}$ is set to 6.  The traffic density is characterized by the number of vehicles per hour per kilometer of roads, which is set to 200, 400, 600 for representing the low, medium, and high traffic density, respectively. The period of information collection at the BS (also the period of running ROPE) is $\tau = 1$ s. 
As for the early warning scheme, the past three-second information ($T = 3$) along with the current information is leveraged for mobility prediction. 
As for the path verification protocol, $\delta = \delta_1 = 100$ ms, $\delta_2 = 70$ ms, and $\delta_3 = 40$ ms.

\begin{figure}[t]
	\centering
	\includegraphics[width=7cm]{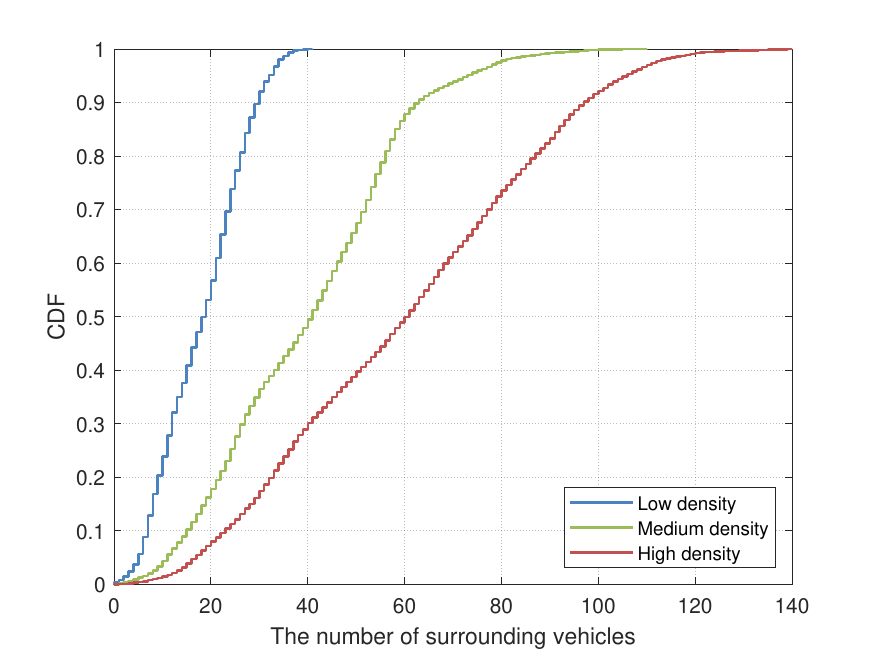}
	\caption{CDF of the number of surrounding vehicles.}
	\label{fig:nvehicle}
\end{figure}

\begin{figure}[t]
	\centering
	\includegraphics[width=7cm]{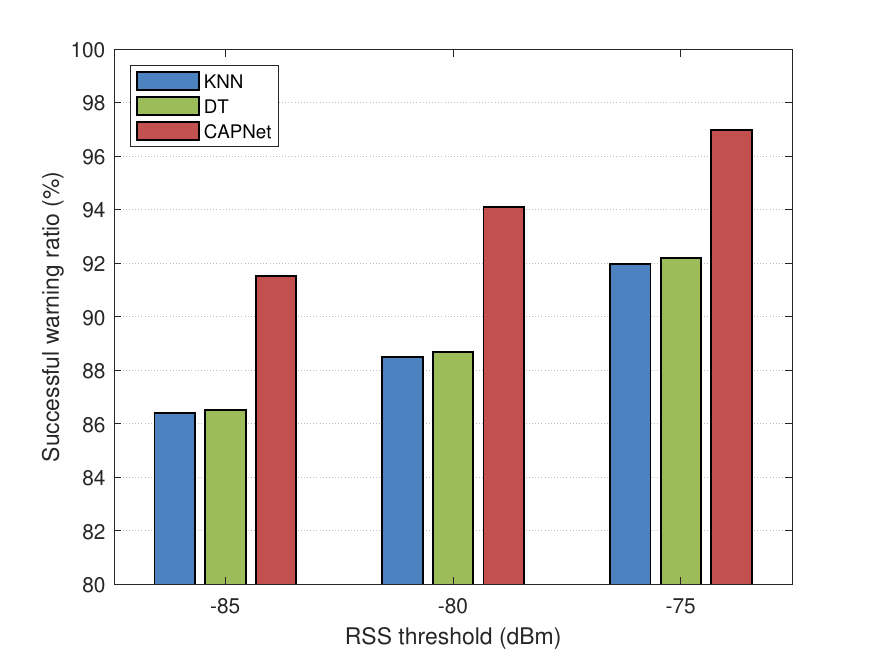}
	\caption{Successful warning ratio under various RSS threshold.}
	\label{fig:warn}
\end{figure}

\subsection{Validation of the Early Warning Scheme} \label{subsec:warn}
In this subsection, early warning scheme is validated by analyzing the impact factors of V2I links and evaluating successful warning ratio of different prediction models. Successful warning ratio is defined as the number of successfully warned V2I links over all tested V2I links.

First, the impact factors of V2I links are analyzed based on the V2X database. To demonstrate the impact of spatial location on V2I links, RSS heatmap in high-mobility scenes is exhibited in Fig.~\ref{subfig:heat}. The simulated area is divided into 5 m $\times$ 5 m grids. For vehicles falling into a grid, their V2I link strength is averaged to represent RSS of this grid. 
One observation is that the grids near the four BSs take on remarkably high RSS, and the grids farther from the BSs take on lower RSS. This is because longer transceiver distance leads to higher path loss and hence lower link strength. 
Another observation is that sharp change in RSS appears on the boundary of some intersections. The reason may be that when vehicles take turns and pass through intersections, their LOS links become NLOS links due to blockage of buildings. Although NLOS links and LOS links with similar transceiver distance have similar path loss, NLOS links suffer prominent link degradation due to the shadowing effect.
To illustrate the impact of traffic density on V2I links, the number of vehicles surrounding the V2I links is calculated according to the $\text{GEMV}^2$ \cite{boban2014geometry,aygun2016geometry} channel model and displayed in Fig.~\ref{fig:nvehicle}.
The cumulative distribution function (CDF) curves show that higher traffic density leads to more surrounding vehicles, so V2I links may encounter more multi-path effect and blockage by vehicles.
Therefore, V2I link strength is influenced by complex factors and the design of CAPNet is feasible in capturing both the explicit factors of spatial location and the implicit factor of traffic density.

Second, successful warning ratio of different prediction models is evaluated in the early warning scheme. Particularly, the database of V2I links is split into a training set, a validation set, and a test set according to the ratio of 6:2:2. All these data sets contain items or instances in low, medium, and high traffic density.
Two models in addition to CAPNet are considered, which are K-nearest neighbors (KNN) and decision trees (DT). Both models are commonly used machine learning methods, which are fed with explicit factors as input features and output deterministic V2I link strength. 
After training and hyper-parameter tuning on the training set and the validation set, these models conduct V2I link inference on the test set. Regarding the test set, successful warning ratio of these models is calculated under different RSS threshold. As shown in Fig.~\ref{fig:warn}, CAPNet obtains the highest successful warning ratio across different RSS threshold, which is significantly better than KNN and DT. The advantage comes from the probabilistic inference ability of CAPNet, which utilizes explicit factors to estimate large-scale strength level and leverages implicit factors to estimate small-scale strength variation.

\begin{table*}[t]	
	\renewcommand{\arraystretch}{1.25}
	\caption{Comparison between the proposed method and its variant method.}	\
	\centering  
	\setlength{\tabcolsep}{14pt}
	\begin{tabular}{c|cc|cc|cc}
		\toprule    
		\multirow{2}{*}{\textbf{Metrics}} & \multicolumn{2}{c|}{\textbf{Low density}} & \multicolumn{2}{c|}{\textbf{Medium density}} & \multicolumn{2}{c}{\textbf{High density}} \\
		& ROPE  & ROPE-  & ROPE  & ROPE- & ROPE  & ROPE- \\ \hline
		$P_S \text{(dBm)}$  &  \underline{-65.60}  &  -65.82 &  \underline{-64.77}  &  -65.22 &  \underline{-66.16}  &  -66.41	\\ 
		$P_C$  &  \underline{1.0000}  &  0.9999 &  \underline{0.9997}  & 0.9994 &  0.99970  &  \underline{0.99972}	\\ 
		$P_H$  &  \underline{2.59}  &  2.72 & \underline{2.72}  &  2.88 &  \underline{2.91}  &  3.05	\\
		$P_Q \text{(\%)}$	 &  \underline{92.15}  &  91.11 &  \underline{91.43}  &  89.96 &  \underline{89.43}  &  88.54	\\ 	
		\bottomrule 
	\end{tabular}
	\label{tab:robust}
\end{table*}

\subsection{Robustness to Link Failures}
Based on the evaluation results in Section~\ref{subsec:warn}, although CAPNet achieves the highest successful warning ratio, it still cannot achieve 100\% prediction accuracy. This implies that imperfect prediction cannot be avoided and predictive routing may suffer link fails, so multi-path selection and path verification are needed to enhance routing robustness.

As shown in Table \ref{tab:robust}, the proposed method is compared with its variant method for validating robustness to link failures. For simplicity, ROPE indicates the complete method that finds three best feasible paths and executes the path verification protocol, while ROPE- indicates the variant method that only finds the best feasible path without the path verification protocol. Both methods apply the same early warning scheme.
When the RSS threshold is set to $\gamma_{th} = -80$ dBm under different traffic density, the table shows average performance of path strength $P_S$, path connectivity $P_C$, path hop count $P_H$, as well as path qualification ratio $P_Q$ of the routing paths. Note that the path strength here and thereafter indicates RSS of the end-to-end path, which is not normalized as in Section~\ref{sec-routing} for more intuitive exhibition.
For low, medium, and high traffic density, better value of each metric is underlined (i.e., higher $P_S$, higher $P_C$, lower $P_H$, and higher $P_Q$). 
It is observed that ROPE- only obtains better path connectivity than ROPE in the high-density scene, while ROPE obtains better path strength, path hop count, and path qualification ratio across all traffic density.
These observations demonstrate superiority of ROPE under all traffic scenes. Instead of directly switching to the predicted path in ROPE-, ROPE conducts link and path check as well as path mending procedures for the predicted three paths, which can test prediction accuracy and quickly handle link failures before switchover. By virtual of these remedy solutions, ROPE obtains superior performance and higher robustness.

\begin{figure*}[t]
	\centering
	\subfigure[Low density] { 
		\label{subfig:rssld}
		\includegraphics[width=0.66\columnwidth]{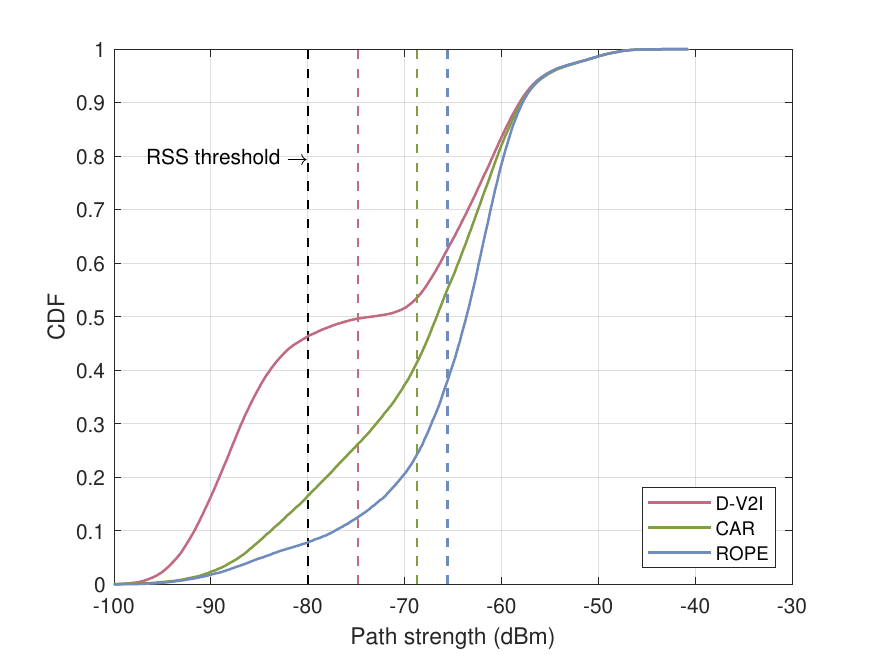}}
	\subfigure[Medium density] { 
		\label{subfig:rssmd}
		\includegraphics[width=0.66\columnwidth]{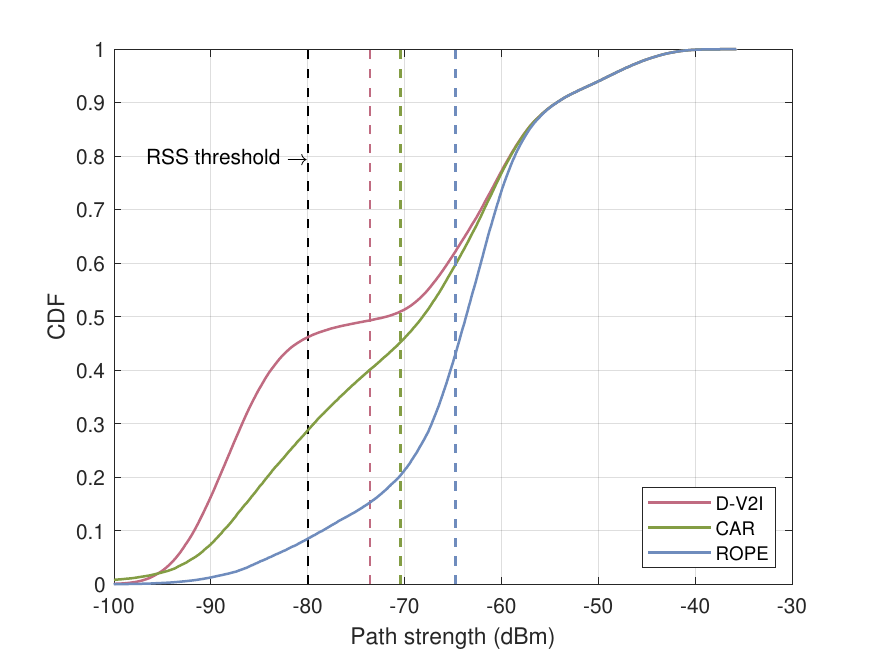}}
	\subfigure[High density] { 
		\label{subfig:rsshd}
		\includegraphics[width=0.66\columnwidth]{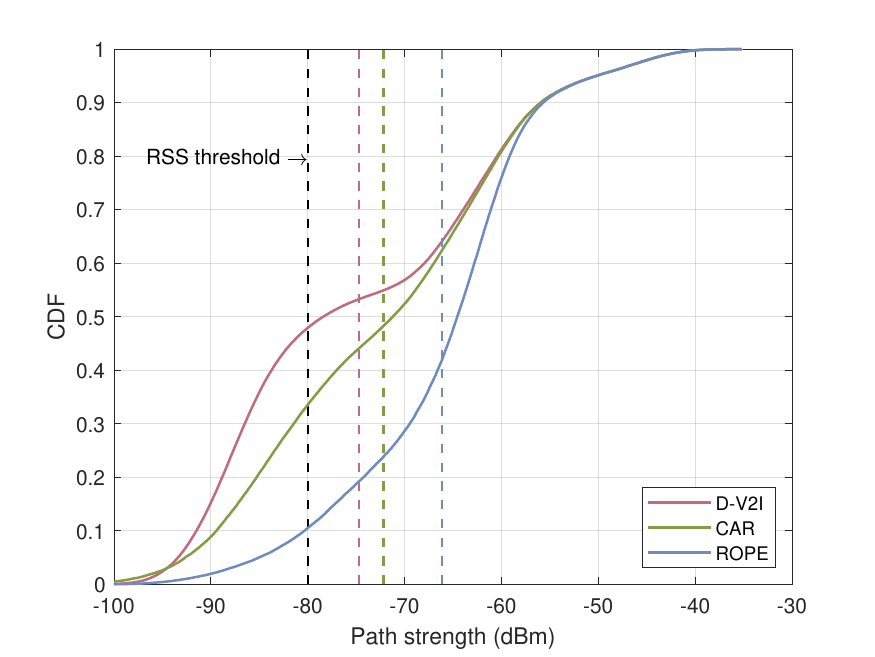}}
	\caption{Cumulative distribution function of path strength under various traffic density.}
	\label{fig:rss}
\end{figure*}

\begin{figure*}[t]
	\centering
	\subfigure[Low density] { 
		\label{subfig:qosld}
		\includegraphics[width=0.66\columnwidth]{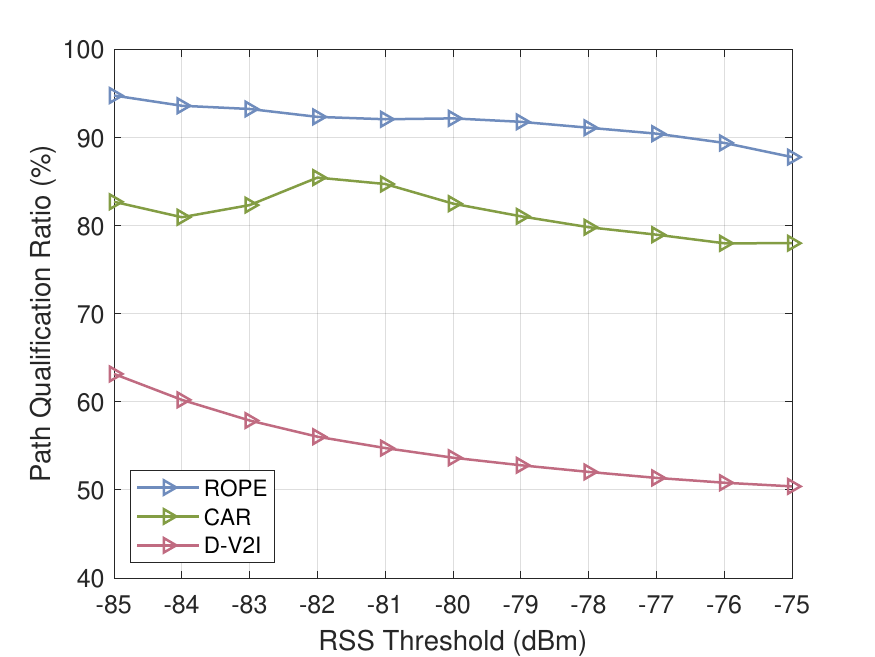}}
	\subfigure[Medium density] { 
		\label{subfig:qosmd}
		\includegraphics[width=0.66\columnwidth]{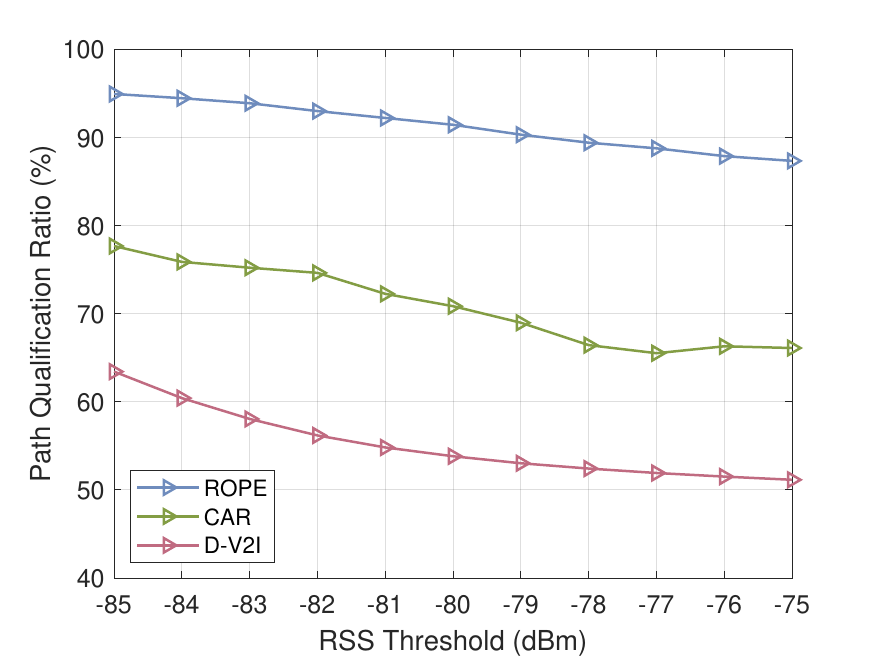}}
	\subfigure[High density] { 
		\label{subfig:qoshd}
		\includegraphics[width=0.66\columnwidth]{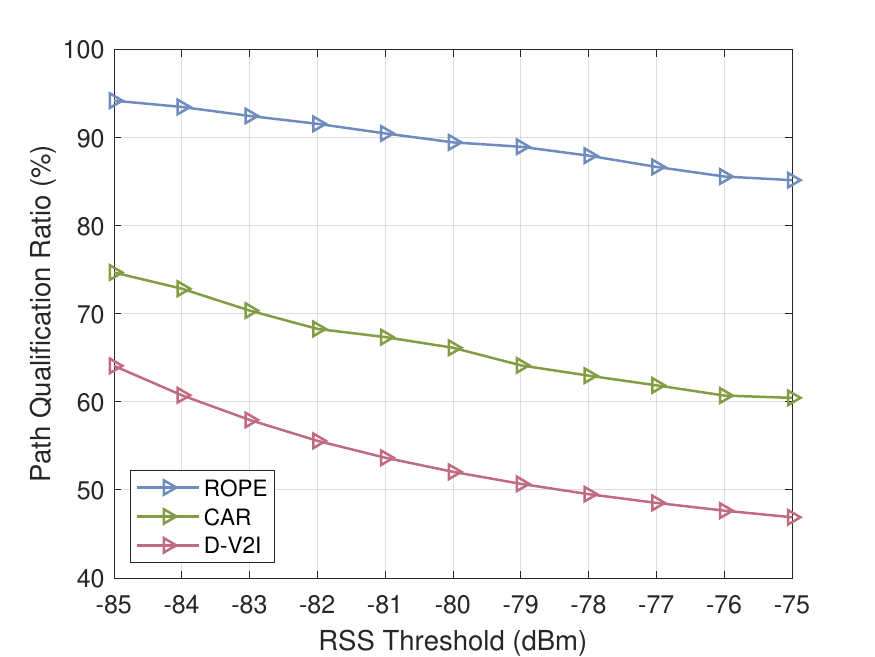}}
	\caption{Path qualification ratio versus RSS threshold under various traffic density.}
	\label{fig:qos}
\end{figure*}

\subsection{Comparison With Baseline Methods}
In this subsection, the proposed ROPE method is compared with baseline methods in terms of path strength and path qualification ratio under various traffic density. Specifically, D-V2I is the baseline using direct V2I links, and CAR~\cite{kumbhar2021novel} is the predictive routing method that selects multi-hop path based on the estimated connection duration.
For fair comparison, CAR is tailored to fit in our framework, where routing is performed on the virtual topology after V2I link inference.

First, cumulative distribution function (CDF) of path strength under RSS threshold $\gamma_{th} = -80$ dBm is illustrated in Fig.~\ref{fig:rss}. 
The dashed line in black depicts the RSS threshold and other dashed lines depict average path RSS of different methods. As for low density, there are around 46.34\%, 16.55\%, and 7.84\% paths determined by D-V2I, CAR, and ROPE that cannot exceed the RSS threshold. The average path RSS of these methods are -74.79 dBm, -68.72 dBm, -65.60 dBm. As for medium density, there are around 46.18\%, 28.77\%, and 8.54\% paths determined by D-V2I, CAR, and ROPE that cannot exceed the RSS threshold. The average path RSS of these methods are -73.60 dBm, -70.45 dBm, -64.77 dBm. As for high density, there are around 47.94\%, 33.58\%, and 10.54\% paths determined by D-V2I, CAR, and ROPE that cannot exceed the RSS threshold. The average path RSS of these methods are -74.71 dBm, -72.18 dBm, -66.16 dBm. 
Based on these results, we can see that ROPE achieves dramatic performance gain over the baseline methods. With multi-hop V2X routing, ROPE reduces V2I link deterioration ratio by 38.50\%, 37.64\% and 37.40\% under low, medium, and high traffic density. With path strength maximized for multi-path selection, ROPE increases average path strength of CAR by 3.12 dB, 5.68 dB, and 6.02 dB under low, medium, and high traffic density.

Second, path qualification ratio versus RSS threshold is illustrated in Fig.~\ref{fig:qos}. As RSS threshold increases, it becomes harder to find qualified paths that satisfy the QoS requirements in terms of path strength, path connectivity, and path hop count. Thereby, path qualification ratio of the three methods display a downward trend. Even so, ROPE still obtains stably high path qualification ratio across different traffic density, which is 94.75\% $\sim$ 87.76\% in low density, 94.93\% $\sim$ 87.33\% in medium density, and 94.17\% $\sim$ 85.14\% in high density. 
The metric of D-V2I degrades to 77.99\% in low density, 65.51\% in medium density, and 60.45\% in high density. These results show that ROPE significantly outperforms the baseline methods, which is benefited by conducting QoS-driven path selection and enhancing robustness to path failures.

\section{Conclusion} \label{sec-conclusion}
In this paper, a robust predictive routing framework (RODA) was proposed for the Internet of Vehicles to fulfill QoS provision of cloud service. 
With novel design of the early warning scheme, the virtual routing mechanism, and the path verification protocol, RODA addressed the challenges of seamless path switchover, QoS-driven path selection, and robustness to link failures. 
Extensive simulations demonstrated the effectiveness of RODA under various RSS threshold and traffic density. 
By detecting V2I link deterioration in advance and finding top3 paths under the QoS requirements, RODA improved path strength significantly and achieved higher path qualification ratio than the baseline methods. By conducting path verification on the top3 paths, RODA enhanced routing robustness to link failures caused by imperfect prediction.
For future work, resource allocation schemes will be developed to collaborate with the proposed routing framework for optimizing packet-level network performance.

\bibliographystyle{IEEEtran}
\bibliography{IEEEabrv,ref.bib}



\end{document}